# Growth of International Collaboration in Science:

# Revisiting Six Specialties

Caroline S. Wagner,[a] Travis A. Whetsell,[a] and Loet Leydesdorff [b]

**Abstract**

International collaboration in science continues to grow at a remarkable rate, but little agreement exists about dynamics of growth and organization at the discipline level. Some suggest that disciplines differ in their collaborative tendencies, reflecting their epistemic culture. This study examines collaborative patterns in six previously studied specialties to add new data and conduct analyses over time. Our findings show that the global network of collaboration continues to add new nations and new participants; each specialty has added many new nations to its lists of collaborating partners since 1990. We also find that the scope of international collaboration is positively related to impact. Network characteristics for the six specialties are notable in that instead of reflecting underlying culture, they tend towards convergence. This observation suggests that the global level may represent next-order dynamics that feed back to the national and local levels (as subsystems) in a complex, networked hierarchy.

[a] The Ohio State University, John Glenn College of Public Affairs, Battelle Center for Science & Technology Policy, Columbus, OH USA, email: wagner.911@osu.edu; travis.whetsell@gmail.com
[b] University of Amsterdam, Amsterdam School of Communication Research (ASCoR), PO Box 15793, 1001 NG Amsterdam, The Netherlands; email: loet@leydesdorff.net

# 1. Introduction

In earlier studies, two of us examined growth in international collaboration in six specialties of science for the period 1990-2005. We showed that growth could not be tied to equipment use or access to materials (Wagner, 2005), but all six cases conformed to a power law distribution, suggesting that preferential attachment (Jeong et al., 2003) could explain the growth of international coauthorship relations (Wagner & Leydesdorff, 2005). This finding accords with research on international collaboration in science as a network, where a pattern of communication enables exchange, interdependent flows of resources, and reciprocity that enhance research processes (Palla et al., 2007; Hoekman et al., 2010). Network analysis has emerged as a meaningful way to study the distinctive form of research communications that transcend institutions and nations (Gilsing et al., 2008; Zeng et al., 2010).

International collaboration is particularly interesting to study because it has grown at a remarkable rate since 1990 (Adams et al., 2005). By 2011, internationally coauthored papers accounted for 25 percent of Web of Science records, up from 10 percent in 1990 (Wagner et al., 2015). Many more nations participate in global collaboration than was the case two decades ago (Bornmann et al., 2015). Some part of the increased activity is tied to participation in large-scale scientific projects such as those taking place at CERN, or the international human genome project. However, "big science" alone cannot explain the growth: many "small science" projects at the international level are based upon the shared interests of otherwise unrelated parties, working independently of organizing imperatives or shared resources, to find reasons to cooperate despite geographic distance.

Some aspects of international collaboration are self-evident: for example, long-distance collaboration is costlier to practitioners in terms of time and treasure. As Barjak and Robinson (2008) point out: "It is clear that international collaboration must bring additional benefits which outweigh the transaction costs; otherwise it would be hard to explain its impressive growth…" (p.25). It appears that the extra effort attracts greater citations: Glänzel and Schubert (2001) showed that international publications have higher-than-expected citation rates in all scientific fields, a finding supported by others (e.g, Narin et al., 1991; Persson et al., 2004; He, 2009).

A second observation is that distance is less of a barrier than it was in the past: Frenken et al. (2009) showed an increase in long-distance collaboration. Collaborative ties have shown a proximity effect (Katz, 1994; Glänzel & Schubert, 2005), but that appears to be diminishing over time (Choi et al., 2015). Territorial borders have become less important to researchers (Hoekman et al., 2010), although there remains considerable heterogeneity between regions in their propensity to cooperate, in part because of differences in investment and capacity. Adams (2013) showed higher growth rates in numbers of publications at the international level. Many have pointed to the role of the Internet in facilitating long distance collaborations (Sonnenwald, 2007).

Gazni et al., (2012) supported Glänzel and de Lange (2002) by showing growth in the number of coauthorships per paper at the international level, as well as growth in the average number of organizations and nations per co-publication. Wuchty et al. (2007) analyzed growth in team sizes, showing growth across all fields of science (national and international). Collaboration is also influenced by increasing mobility of researchers (Jonkers & Cruz-Castro, 2013), with many researchers spending time in different places (Halevi et al., 2016) and creating new connections in ways that become a link in a network to be activated when the needs of research dictate a new direction. The mobility factor, the increased numbers of researchers, and complexity of problems are some of the reasons that publications show increasing numbers of coauthors.

It is widely supported that collaborative activity differs by field. The literature is mixed on distinguishing features of collaboration by field (international or team-based). Some divergent findings can be attributed to differences in data collection and analytical methods (as well as differences in what is counted as a "field" or "discipline"). Wuchty et al. (2007) examined many fields over decades: They showed growth in team sizes by discipline of as much as 135 percent in some fields, with the highest growth in Physics, Environmental Science, and Medicine. Mattsson et al. (2008), focusing on international collaboration, found higher collaboration rates in Physical, Chemical, and Earth Sciences, and for Life Sciences. Mattsson et al. (2008) did not confirm Wuchty et al.'s finding for higher rates for Medicine (most likely due to a difference in team versus international approach to the data).

Abt (2007), studying just one year (2005) suggests that simultaneity of data access influences the tendency to collaborate – in other words, if scientists from many countries have access to specific data at the same time, international collaborative papers are more likely to emerge. Moreover, Abt (2007) showed that, for 2005, fields with the largest numbers of researchers working internationally had the highest impact factors, particularly in Medicine (with a focus on the specialties of Cardiology and Neurosciences). Abt (2007) found the fields with the highest percentage of internationally coauthored papers for 2005 were Astronomy, Physics, Geophysics, and Biology; the less populated fields were Mathematics, Engineering, and Geosciences.

Newman (2001b) analyzed data over three years and showed that, in purely theoretical disciplines within Physics, the average number of collaborators per paper is lower compared to experimental fields, which would support Abt's (2007) theory about access to data being an important factor in the production of coauthored papers. Gazni et al. (2012) showed that collaboration tends to be highest in the Life Sciences, which could support Wuchty's findings about Medicine and Biology. Several studies have shown Mathematics as among the least collaborative fields (Newman (2001b). Glänzel & de Lange (2002) also showed that the median number of collaborating countries is relatively high in Physics, and low in Mathematics.

This paper does not seek to settle the question of which fields are higher or lower in international collaboration: the literature tends to support the assessment that Physics, Biology, and Environmental (Geo) Sciences have larger teams, more papers, and higher impact, while Engineering, Agriculture, and Mathematics have smaller team sizes, fewer papers, and lower impact.

With this in mind, this study examines six specialties embedded within the six broader fields listed in parentheses: 1) Astrophysics (Physics); 2) Virology (Biology/Medicine); 3) Seismology (Environmental/Geo); 4) Polymers (Engineering); 5) Soil Science (Agriculture); and 6) Mathematical Logic (Mathematics) to add new data and analysis. In addition, we compare analysis of these data with an analysis of all publications in all fields in the years studied. Our interest is in exploring the dynamics of knowledge creation at the international level. We focus at the specialty level to test whether networked sub-communities reveal distinctive topological features, perhaps based upon the needs of the specialized research. Further, we test whether distinct specialties showed different growth rates that might indicate their relationship to the global system as a whole. Finally, we have a longer term interest in exploring the make-up of the global network of collaboration to ask whether this network is evolving into a next-order dynamic that feeds back to the national and local levels.

Connections are explored at the nation-to-nation level with the assumption that nations represent an underlying political and cultural structure of scientific support. We address the following questions: Do disciplines of science exhibit significant differences in collaborative groupings? Does the scope of international collaboration influence quality (expressed as citations)? Can we derive significance from differences among the disciplines that can shed light on their epistemic cultures? Based on the literature review, we expect to find that the resulting measures will differ by discipline with Astrophysics, Virology, and Seismology exhibiting the highest collaborative tendencies, and with Soil Science, Polymers, and Mathematical Logic exhibiting lower. We expect to find diminishing returns to team size at some critical point. On the basis of earlier findings (Newman, 2001a), we expect to see growth at the network level showing denser, more connected, and more equitable networks.

## 2. Data and Methods

The nation-to-nation links for all coauthored papers were provided by Elsevier for 2008 and 2013. In addition, Elsevier collected all papers from specified journals at the specialty level to match earlier analysis (see Appendix A) (Wagner, 2005). Elsevier provided the metadata on numbers of publications, numbers of countries per paper, and citation impact from these journals for six specialties for 2008 and 2013. The patterns for six specialties are analyzed in terms of numbers, networks, and regressions to examine the collaborative structure of the fields. The whole networks for 2008 and 2013 are presented as a baseline comparison for the six cases. The regression tests are made on all subjects catalogued by Elsevier, and the six specialties. Earlier data for the same six specialties for 1990 and 2000 were drawn from the same journals from the Web of Science (Wagner, 2005) and are compared to the newer findings.

We reconstructed the nation-to-nation collaborations into adjacency matrixes $\mathbf{X}_{ijt}$, where $i$ and $j$ represent distinct nations, and $t$ represents the year 2008 or 2013. Next, we calculated cosine-normalized weights based on number of publications with a given nation-nation combination. The matrices were analyzed using UCINet (Borgatti et al., 2002) and Pajek, (and double-checked in Gephi) to derive statistical properties of the structure of the networks of each subfield in both 2008 and 2013 and enable comparisons to similar data from 1990 and 2000.

We employ several measures to analyze the structural properties of the networks. The number of nodes, edges, and diameter are measures of size. Average degree and density are measures of structural cohesion. Clustering and betweenness measure cliquishness, small worlds, and properties of sub-group redundancy to examine openness. Diameter measures how many edges (steps) are necessary to "step" from any node to reach any other node in the network (De Nooy et al. 2011; Monge & Contractor 2003) to reveal local search capacity.

Next we use mixed effects regression to estimate the marginal impact of each additional nation added to collaborations on field weighted citation impact (FWCI). The mixed model was chosen in order to model the specific country combination as a random effect and the year as a fixed effect. This analysis is conducted separately for each of six specialties, and for an "all fields" category. The FWCI is calculated by Elsevier; the calculation requires that the program have access to full data of all publications and citations for each discipline by year (see also, Halevi et al., 2016). By comparing the citations for the specialty to the average number of citations per paper across the entire discipline, an index of the attention garnered by a subfield is shown. We estimate separate models for each of six subfields and one model for all of science (that is, all publications in the database in that year) combined.

In contrast to many prior studies of citation impact, we specifically analyze the collaboration itself, i.e. the nation-to-nation instance or combination, and not the publication. Each observation lists the total number of publications with a given country-country combination. (Data are available on figshare.) FWCI is used to represent the impact of the particular nation-to-nation combination within the subfield. Other variables included in these models are discussed further in the regression results section.

3. **Results**

The results are presented in three sections. First, the results of the overall counts, showing new data and then comparing it to previously collected data are presented. Second, network analyses are discussed and compared. Third, the results of the mixed effects regression analysis are presented.

*3,1. International Collaborative Tendencies*

Table 1 shows the counts for the six specialties and the baseline of fields (all scientific publications from the referenced year listed as All-Fields) in 2008 and 2013 for Elsevier data. The number of nodes (nations) represented in the international collaboration network for All-Fields appears very high (compared to the United Nations count of 193 countries) because many small nations such as East Timor and even North Korea are recorded as addresses in the database. We expected to find Astrophysics (Physics) at the top of the list of most internationalized specialties, but Virology (Biology/Medicine) is the most internationalized specialty of the sample, with 120 nations participating in international collaboration in 2013. As expected, Mathematical Logic (Mathematics) is the least internationalized, with just 32 nations participating in international collaborations in 2013, which decreases by 3 from 35 in 2008. As expected, Seismology (Environmental/Geo science) has a large number of nations participating in international collaboration. Against expectations, Soil Science (Agriculture) is highly internationalized in these years, just behind Seismology and Virology as the sciences with the most nations participating in collaboration. Astrophysics (which we expected to be high) and Polymer Science (Engineering) (which we expected to be low) fall into the middle in terms of numbers of nations participating in international collaborations in these years.

Table 1. Counts of collaborative activities at the international level for six specialties, 2008 and 2013, raw data from Scopus, see technical note #2 for discussion

| Field | Year | Nodes (nations) | Edges (links) |
|---|---|---|---|
| All-Fields | 2008 | 228 | 3346 |
|  | 2013 | 230 | 4230 |
| Astrophysics | 2008 | 81 | 936 |
|  | 2013 | 87 | 1251 |
| Mathematical Logic | 2008 | 35 | 74 |
|  | 2013 | 32 | 58 |
| Polymer Science | 2008 | 75 | 334 |
|  | 2013 | 72 | 391 |
| Seismology | 2008 | 93 | 466 |
|  | 2013 | 101 | 619 |
| Soil Science | 2008 | 92 | 373 |
|  | 2013 | 100 | 429 |
| Virology | 2008 | 112 | 611 |
|  | 2013 | 120 | 693 |

The number of edges (links) among the participating nations tells a slightly different story to counts of participating nations. Counts of edges are closer to expectations (Table 1) in that Astrophysics has the largest number of connections at the global level, with more than 1000 edges in 2013. Virology follows, with 693 edges across nations in that year. Mathematical Logic is the least internationally connected among the sets, with 58 edges in 2013, down from 74 links in 2008.

Table 2 shows the number of participating nations and the number of international collaborations for each specialty for the four years studied (please see technical note 2 for additional details). Looking back over time to earlier analysis, we see that the number of nodes (nations) and edges (links among nations) for all fields has risen rapidly from 1990. All fields show strong growth in the number of nations with researchers working together. All six specialties added between 18 and 60 new nations to the list of collaborating partners, as can be seen in Table 2. Of the six cases, Mathematical Logic, Seismology, Soil Science, and Virology all show near or more than doubling of the number of nations participating in global collaborations between 1990 and 2013. All fields combined saw more than a 120 percent increase in the amount of connectivity revealed in the data.

Table 2. Network size measures for all publications and for six specialties and changes between first and last year, 1990, 2000, 2008, and 2013.

| Net Measure | 1990 | 2000 | 2008 | 2013 | Change between 1990 and 2013 |
|---|---|---|---|---|---|
| Nodes (Nations) | 172 | 192 | 228 | 230 | 58 |
| Edges (Links) | 1926 | 3537 | 3346 | 4230 | 120% |
| Diameter (Steps) | 3 | 3 | 3 | 3 | no change |
| Nodes | 50 | 73 | 81 | 87 | 37 |
| Edges | 337 | 745 | 936 | 1251 | 271% |
| Diameter | 4 | 3 | 4 | 3 | Decrease |
| Nodes | 14 | 33 | 35 | 32 | 18 |
| Edges | 17 | 75 | 74 | 58 | 241% |
| Diameter | 5 | 5 | 5 | 5 | no change |
| Nodes | 50 | 71 | 75 | 72 | 22 |
| Edges | 110 | 311 | 334 | 391 | 255% |
| Diameter | 5 | 5 | 4 | 4 | Decrease |
| Nodes | 53 | 80 | 93 | 101 | 48 |
| Edges | 166 | 440 | 466 | 619 | 273% |
| Diameter | 5 | 4 | 4 | 3 | Decrease |
| Nodes | 40 | 80 | 92 | 100 | 60 |
| Edges | 66 | 247 | 373 | 429 | 550% |
| Diameter | 5 | 5 | 4 | 4 | Decrease |
| Nodes | 60 | 89 | 112 | 120 | 60 |
| Edges | 190 | 338 | 611 | 693 | 265% |
| Diameter | 4 | 4 | 4 | 4 | no change |
| | Data source: Web of Science | Data source: Scopus | | | |

Against expectations, Soil Science grew the fastest of the fields with a 550 percent change increase in edges (links) between 1990 and 2013, growing much faster than all fields combined (the global whole). This is unexpected because Agriculture tends to be much less internationalized than other fields. The number of nations connected through Soil Science research more than doubles from 40 to 100 nations supporting collaborations. While Soil Science shows the most impressive growth, all the specialties grow in number of edges (links) by more than 200 percent between 1990 and 2013.

In a communication structure, it is useful to know, not just who is connected to who, but how indirect ties can enable knowledge flow or search from one node to another. Distance/diameter measures the number of steps or intermediaries needed for movement from one node to another: the shorter the distance, the easier it is to exchange information or find new partners. For all the data sets, Table 2 shows the diameter is between 3 and 5—an amazingly low number--that suggests that the global networks are tightly linked together.

*3,2. Network measures*

Network analysis provides measures that give insight into the nature of systems of communications within disciplines. We apply these measures to the networks built from nodes (nations) and edges (links). The network is assumed to be undirected, in that information can be exchanged reciprocally between two nodes (as opposed to directed, where information is passed in one direction only). The average degree measures the extent to which possible connections are actually realized in the network. (The degree $\kappa$ of a node is the number of edges connected to it. The average degree measures these across the network for all nodes.) Density measures the extent of connections made out of all that maximally be made. Betweenness (centralization) measures the role of hubs in determining structure and flow. Clustering measures cliques within the network.

The step-wise change in the number of nodes in the network (represented in the number of nations) combined with a substantial increase in the number of edges (links), contributes to the significant increase in the structural cohesion of the network as measured by the average degree of the network, which grows from 22.4 in 1990 to 73.6 in 2013, shown in Table 3. This observation of structural cohesion is further supported by the density measure which suggests extensive growth of connectivity among nations from .13 to .30 in 2013. The data show that all subjects (the global network) is highly interconnected (dense), with all nations connected to all other nations by some path (probably many paths) of intermediate connections, although not all nations are present in each of the specialties (Table 2).

Table 3. Network statistics for All Fields and for Six Specialties, 1990, 2000, 2008, 2013

| Scientific Field | Net Measure | 1990 | 2000 | 2008 | 2013 |
| --- | --- | --- | --- | --- | --- |
| All Fields | Avg. Degree | 22.40 | 36.90 | 58.70 | 73.60 |
| | Density | 0.13 | 0.19 | 0.26 | 0.30 |
| | Betweeness | 0.26 | 0.16 | 0.11 | 0.08 |
| | Clustering | 0.78 | 0.79 | 0.61 | 0.80 |
| Astrophysics | Avg. Degree | 13.48 | 20.41 | 23.11 | 28.76 |
| | Density | 0.28 | 0.28 | 0.29 | 0.33 |
| | Betweeness | 0.12 | 0.12 | 0.10 | 0.09 |
| | Clustering | 0.59 | 0.64 | 0.67 | 0.68 |
| Mathematical Logic | Avg. Degree | 2.43 | 4.55 | 4.23 | 3.63 |
| | Density | 0.19 | 0.14 | 0.12 | 0.12 |
| | Betweeness | 0.31 | 0.39 | 0.32 | 0.27 |
| | Avg. Cluster | 0.38 | 0.28 | 0.26 | 0.29 |
| Polymer Science | Avg. Degree | 4.40 | 8.76 | 8.90 | 10.86 |
| | Density | 0.09 | 0.13 | 0.12 | 0.15 |
| | Betweeness | 0.43 | 0.23 | 0.27 | 0.17 |
| | Clustering | 0.25 | 0.38 | 0.39 | 0.42 |
| Seismology | Avg. Degree | 6.26 | 11.00 | 10.02 | 12.26 |
| | Density | 0.12 | 0.14 | 0.11 | 0.12 |
| | Betweeness | 0.33 | 0.24 | 0.31 | 0.30 |
| | Clustering | 0.37 | 0.47 | 0.40 | 0.46 |
| Soil Science | Avg. Degree | 3.30 | 6.18 | 8.08 | 8.58 |
| | Density | 0.08 | 0.08 | 0.08 | 0.09 |
| | Betweeness | 0.63 | 0.24 | 0.23 | 0.17 |
| | Clustering | 0.19 | 0.29 | 0.37 | 0.36 |
| Virology | Avg. Degree | 6.33 | 7.59 | 10.91 | 11.50 |
| | Density | 0.11 | 0.09 | 0.10 | 0.10 |
| | Betweeness | 0.45 | 0.34 | 0.29 | 0.29 |
| | Clustering | 0.36 | 0.30 | 0.35 | 0.35 |
| | | Data source: Web of Science | | Data source: Scopus | |

Table 3 shows the average degree for all subjects and then for the six specialties in 2008 and 2013. Average degree is a useful measure because it shows the extent of connectivity and growth among the relatively fixed number of nations: the number of nations increases step-wise from year to year, but the number of edges (links) grows more quickly as more researchers develop international collaborations contributing to the growth in average degree. Among the specialties, Astrophysics shows the highest degree of the six fields, suggesting that it is the most intensely and redundantly connected at the global level of the specialties. Average degree increases in all six cases.

Density shows variation across the fields, with Polymer Science showing unexpected growth in density, since it was not expected to be highly internationalized, and its research generally does not require sharing of resources or large scale equipment. Apparently, as new nations join the global network in this field, multiple connections are made. Astrophysics also shows strong growth in density, as expected. Clustering is highest for Astrophysics, suggesting that if two nodes are connected, they are likely to be connected to a third partner, as well. (This may be due to articles with many authors.) Mathematical Logic again is the least likely to have dense networks, cliques, or connections, which reflects articles with fewer authors. Figures 1 and 2 show the networks in this specialty.

Betweenness centralization is the variation in the node betweenness centrality divided by the maximum degree variation possible in a network of the same size (DeNooy et al, 2011). Centralization declined in all six cases. This indicates that central nodes are generally becoming less critical to the overall structural cohesion of the international collaboration network. In other words, the larger, scientifically advanced nations are not dominating activity. In Astrophysics, the United States remains dominant in the network—most likely due to the scale of equipment costs. But, in other fields, the larger nations are not retaining positions as strong hubs. In other words, the networks are dense but not highly centralized, thus there are many redundant connections among nations. This may provide multiple opportunities for practitioners to connect to new collaborators. Power is not concentrating in a few large nodes.

Clustering is a measure of the fraction of transitive triples—that is, the probability that nation A, nation B, and nation C are completely interconnected. (Consider that Portugal, Brazil, and Mozambique might be fully connected through soil research, for example.) This measure, taken across the entire network, reveals the extent of interconnectedness and it shows how easily information can travel across the network. At the nation-nation level, the possibility for knowledge exchange and sharing is high, shown in the clustering coefficient that began at a high level in earlier studies, and has grown slightly between 2008 and 2013.

The clustering coefficient is one of the parameters used to characterize the topology of complex networks. We wished to compare these over time, since other research has shown that social networks have higher clustering than random networks (Newman, 2001a). Clustering coefficients show growth over time for four of the six specialties: Mathematical Logic drops (because of its small size), and Virology holds steady at about .35, perhaps because groups of nations work together on specific diseases, and do not cross over much from one to the other grouping. Astrophysics, Mathematical Logic, Polymers, Seismology, and Soil Science, all show growth in clustering coefficients over time, suggesting that participants in the specialties are finding new partnerships among existing members.

In the global network, two phenomena are occurring: the number of spatial neighbors is expanding linearly towards a maximum possible number (number of nations), and the connectivity among nodes is growing. Nodal connectivity is not limited by geography, since the number of researchers within a nation can grow. The average degree—the counts of connectivity per node averaged over the network--for global collaboration increases between 2008 and 2013; these measures are considerably higher than measures made on similar networks for 1990 and 2000 suggesting that many more researchers are linked than was the case in the 1990 or 2000. The increase supports Newman's finding (2001b) of percolation transition, which suggests that participants have established many more connections over time, and they have been joined by new actors (both new nations and new participants from existing nations) who are also making connections, perhaps facilitated by existing connections of those already in the network. This observation is supported by the low levels of clustering shown in Table 3, where we see that cliques are not growing rapidly.

Comparing the findings from 2008 and 2013 (data from Scopus) to a similar analysis in 1990 and 2000 (data from Web of Science), Table 3 shows that the network of collaborating countries in only two of the six the specialties have grown denser over time (which means more connections among existing nations, and new connections from new nations): Astrophysics and Polymers. Four fields remain about the same in density, against expectations and in contrast to the global measure, probably because many new nations joined international collaborations.

To reveal the nature of the underlying network, we visualize the Mathematical Logic networks from 1990 and 2013 in Figures 1 and 2 to show the change in structure and complexity over time. Mathematical Logic is an outlier in the data, perhaps because of a small number of nations in the set. As can be seen in Figure 1, Mathematical Logic collaboration is dominated in 1990 by the United States, which is highly between other nations. This affords U.S. participants a privileged role in disseminating information and connecting people. In Figure 2, the network has grown considerably, and we can see three distinct hubs in the network, with the United States, Great Britain, and China holding central positions. Belgium and Germany also have multiple connections and cohere parts of the network that might spin-off if they were not connected through these lesser hubs. These figures reveal the dynamic behind the drop in betweenness, since by 2013, the United States is no longer as "between" other nations as it was in 1990. The drop in betweenness is notable because it occurs across all specialties and across the global network. It is also possible to see from these figures why density might drop as new entrants arrive, since entrants make one or two connections in this field. All the networks can be viewed on figshare.

Figure 1. The network of collaborative links in Mathematical Logic among nations, 1990 Data: Web of Science

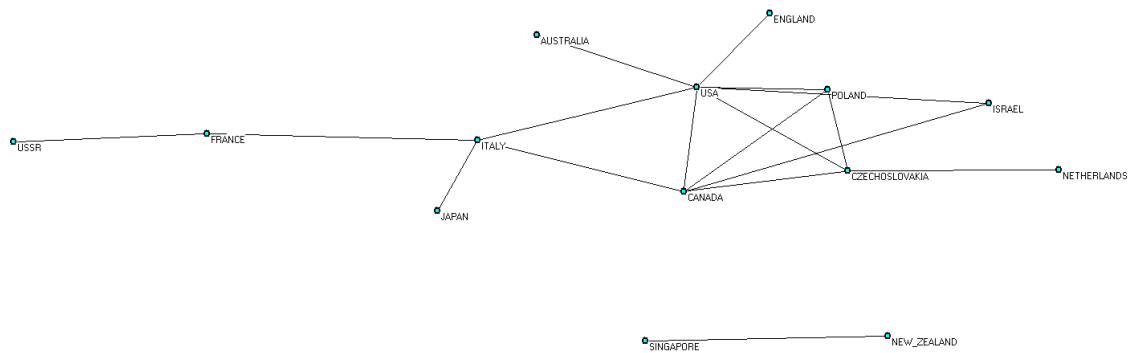

Figure 2. The network of collaborative links in Mathematical Logic among nations, 2013 Data: Sco

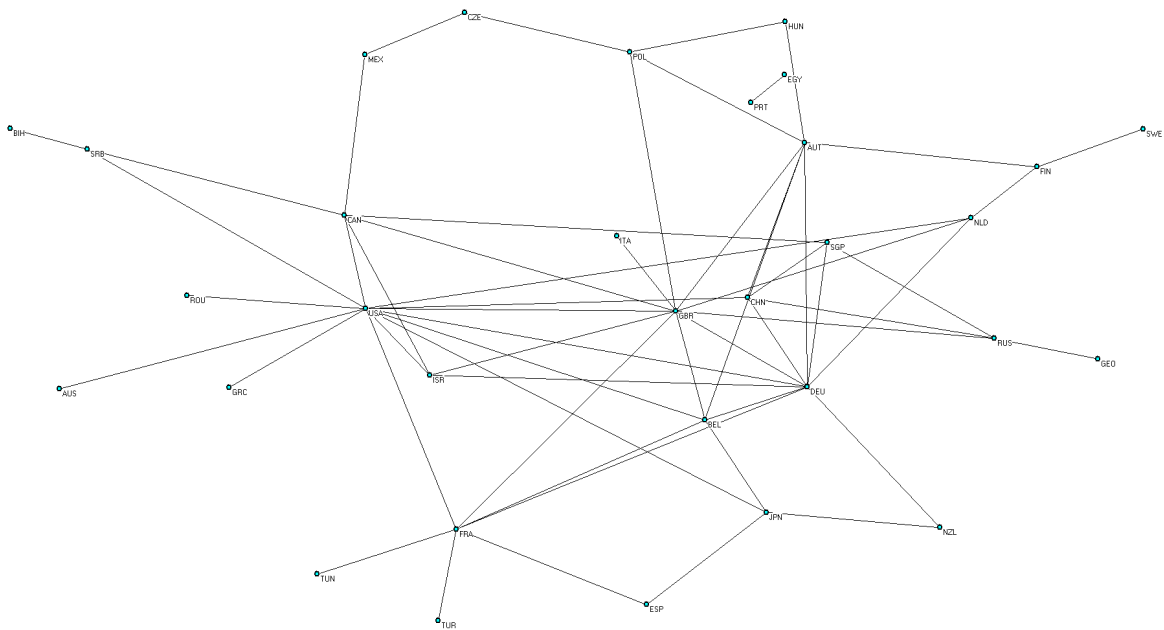

participating, but that also continues to grow in number of new participants from within nations joining the network.

These numbers reveal some distinctive features of the global network. One is that the global network is open to new entrants both at the nation level and at the participant level. Many more nations are represented in the global network over time, and by count, we can say that there are even more new participants from within existing nations who have joined global collaborations. At the broadest level (for all sciences), most nations are within two or three steps from most other nations. At the specialty level, most nations are within four or five steps of others nations (see also the figures to envision the possibility of moving from one node to another in a few steps). Moreover, a theory of adjacency might also suggest that if two nations have participants working together, others from within that nation can gain access through them to global collaborators.

A second notable finding about the global network is that power is being diffused over across the network over time. While large, scientifically advanced nations dominate in terms of numbers, and in terms of elite articles, the networks are not clustering around these nations. This is well illustrated in the Mathematical Logic networks in Figures 1 and 2. As new entrants enter the network beyond 1990, new connections were made among participants with the result being that power and influence is being diffused across the networks. By 2013, there are many nodes with different connections, creating an opportunity for new entrants to come into the network from a much larger number of participants. Centrality drops because several nations now serve as hubs by 2013. We found for all the specialties that large, scientifically advanced countries are less likely to dominate the networks, and they are less and less in a position to impede knowledge flows or block new entrants.

We expected to find differences among the six disciplines, and this is confirmed: The networks revealed for each of the six specialties show variations among them. Table 1 shows that Astrophysics has the highest density measure. This may be an artefact of some articles listing many authors from many countries. The measure suggests that Astrophysics is highly connected across the globe, an interpretation that is supported by the fact that centralization drops over time for that field. As percolation theory suggests, average degree rises because existing nodes (nations) have made new (additive) connections. Of the 230 countries that could possibly be included, 87 are participating in international Astrophysics collaborations in 2013. However, Astrophysics is not the most internationally connected of the fields: Virology is the most internationally interconnected of the specialties.

*3,3. Regression Models*

The regression models are used here to test for the relationship between the scope of the international collaboration and the impact of the collaboration. The analysis tests for the relationship using the number of countries listed in the collaboration for scope and the impact of the collaboration using field weighted citation index (FWCI). The unit of observation in this analysis is the international collaboration. Each observation in the sample is an aggregate of all papers with a particular combination of nations.

Table 4 lists the results of the mixed effects regression analysis. Seven models were analyzed, one for each specialty of interest, and one model for all-fields. The all-fields model includes all international collaborations in Elsevier's Scopus collection for the two years. FWCI is the dependent variable. FWCI is very right skewed, and so the natural log of FWCI plus 0.1 is used, where 0.1 was first added to account for zero values in FWCI. The key independent variable of interest is country count, which simply captures the number of countries listed in a given country combination row. Our data allowed for the inclusion of control variables, such as the number of publications associated with the particular country combination and the year of the collaboration (2008 or 2013). A unique identifier was created for each country combination. This variable is used as a random effect in the mixed effects models below.

Table 4. Results of mixed effects regression models for collaboration in six specialties using the Field Weighted Citation Index

Mixed Effects Regression Results, Dependent Variable = Log of Field Weighted Citation Impact

|  | Polymers | Math | Virology | Soil Science | Seismology | Astrophysics | All Fields |
|---|---|---|---|---|---|---|---|
| Intercept | 241.7*** | 318.17*** | 128.65*** | 175.85*** | 278.8*** | 90.49*** | 167.48*** |
|  | (37.28) | (113.25) | (27.55) | (40.77) | (31.94) | (13.56) | (2.36) |
| Country Count | 0.036 | 0.093 | 0.175*** | 0.16** | 0.153*** | 0.099*** | 0.139*** |
|  | (0.079) | (0.308) | (0.03) | (0.053) | (0.036) | (0.008) | (0.003) |
| Publication Count | 0.0623*** | 0.49 | 0.025*** | 0.076** | 0.041*** | 0.0149*** | 0.0007*** |
|  | (0.015) | (0.23) | (0.007) | (0.024) | (0.010) | (0.0028) | (0.00007) |
| Year | -0.12*** | -0.16* | -0.064*** | -0.088*** | -0.139*** | -0.045*** | -0.084*** |
|  | (0.019) | (0.056) | (0.014) | (0.020) | (0.016) | (0.0067) | (0.0011) |
| Random Effect | 0.0501 | 0 | 0.092 | 0.22 | 0.16 | 0.49 | 1.140 |
| Residual | 1.499 | 2.0967 | 1.0868 | 1.4696 | 1.4488 | 0.727 | 1.239 |
| AIC | 2378.7 | 395.7 | 2978.3 | 2147.6 | 3361.3 | 11325.5 | 677128.4 |
| N | 720 | 109 | 984 | 633 | 1008 | 3752 | 202824 |

Note- first six models were calculated using SAS. R was used for the final model.
Standard Errors in Parentheses
* $p<0.05$, **<0.01, *** $p<0.001$

The results show that the number of countries affiliated with a collaboration is significant for four of the six specialties. Virology has the strongest effect size, followed by Soil Science, Seismology, and Astrophysics. This accords with our expectations based on the hypothesis that international collaboration attracts greater attention, and more nations attract new actors and more attention. Country count maintains a significant effect on citation impact for all fields combined and for four of the six specialties.

The addition of nations is not significant for two fields: Polymers and Mathematical Logic. This may be because these fields do not have reasons to assemble larger teams. Barjak and Robinson (2008) suggest that smaller teams are more effective at creating notable findings, and where equipment or data are not involved, adding more nations might reduce efficiency.

The number of publications is significant for all the models, except for mathematical logic. This suggests that more frequent contact between authors from particular combinations of countries may increase the efficiency of the collaboration, resulting in greater citation impact. Finally, the random effect and residual estimates indicate that the specific country combination explains more of the variation in citation impact for the all-subjects model and the astrophysics model which may be an artefact of the initial dominance of larger, scientifically advanced nations. Conversely, the specific country combination appears to explain very little of the variance in citation impact for the remaining models and none of the variance in the mathematical logic model.

**4. Discussion**

International collaboration in science continues to grow as a whole and within the six specialties studied. The growth reflects the fact that as more nations have capacity to enter the global network, they do so. Supporting expectations, all specialties are more connected over time. Astrophysics, Virology, and Seismology fit expectations for highest collaborative tendencies with higher average degrees than the other three specialties. However, Polymer Science is very similar in average degree to Virology, against expectations, suggesting strong increase in its connectivity. Soil Science increased as well, although more slowly, in average degree. Mathematical Logic meets expectations with lower global participation.

Supporting expectations of a more equitable network is the drop in betweeness centralization in all specialties. This indicates that new entrants are not necessarily entering the network by connecting to the largest hubs. They may be entering based upon regional proximity or similar economic levels and challenges. The drop in betweenness centralization indicates that the network is less dominated by scientifically advanced countries.

Against expectations, the network measures suggest that the specialties appear to be converging towards similar levels of international activity, as shown in Figure 3, where the number of nations participating are shown as growing towards a saturation level. We expected to see distinct community patterns reflective of epistemic cultures, as shown perhaps in the outlier specialty of Mathematical Logic (Figure 1 and 2). Figure 3 shows the convergence of each specialty as a percentage of the total number of nodes in 2013. In other words, while Astrophysics collaborations share equipment and large-scale data sets in a top down-centralized mode, and Virology focuses on many difference diseases in different places in a bottom-up decentralized mode, both fields show similar network structures and rapid international growth. Perhaps more telling is the rise in network participation of Soil Science and Polymer Science which we did not expect to see, but where network measures show similar patterns to the more internationalized sets.

Figure 3. Number of nations as a share of collaborative links, 1990 to 2013

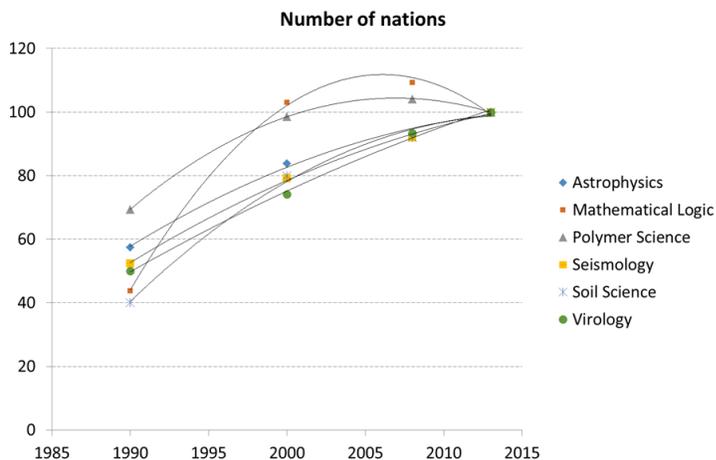

It is possible to suggest that the specialties have taken on properties related to networked communications, rather than unique properties of epistemic cultures. This suggests that the global network has a culture, pathways, and norms of communication specific to its structure, and diverging from national, regional, or disciplinary norms. To examine this further, we combined all specialties, shown in Figure 4, and found support for convergence at the global level.

Figure 4. Convergence of all factors among collaborative links, 1990 to 2013

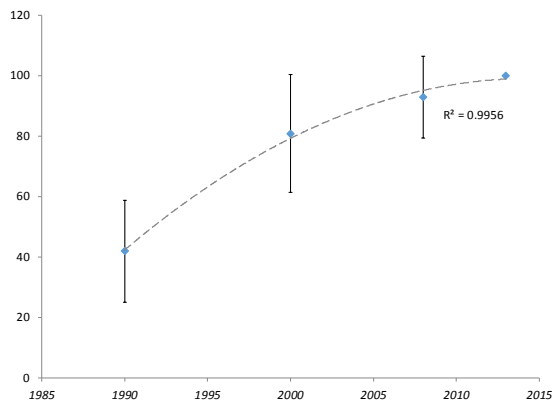

This convergence causes us to speculate that the global network may be developing into an emergent hierarchy similar to that suggested by Simon (1962), who noted that complexity often takes the form of hierarchy. The hierarchy "disciplines" the structure and influences actors. In discussing organization dynamics as an "architecture of complexity," Simon (1962) suggested that "charts of social interactions, of who talks to whom, the clusters of dense interaction in the chart will identify a rather well-defined hierarchic structure…" (p. 469). A complex system, in Simon's view, is one made up of large number of parts that interact in non-simple ways. This definition could fit the networks within the six cases. The networks created for the six cases are indeed charts of social interactions, of who talks to whom, as suggested by Simon. Simon expected that complex systems of these types are made up of levels, towards hierarchies, consisting of successive sets of subsystems, a suggestion which can also be descriptive of the six cases (as well as the global network). For example, Soil Science is nested within Agriculture, which in turn is nested within All Fields. However, a specialty such as Mathematical Logic (which may have more of a transdisciplinary constitution) may be crossed with Philosophy, which could explain its outlier status.

Some parts of Simon's theory of complex architecture appear to apply to the six cases. For example, Simon suggests that hierarchic systems are usually composed of only a limited set of different kinds of subsystems, in various combinations and arrangements. This can be applied to the six cases, since there are only a few subsystems operating within them, such as academic and research institutions, or funding institutions that provide a selection function and feedback role. Schools of thought within disciplines can be characterized as subsystems within the global system. However, Simon also expected that subsystems would be organized in levels or layers or a collection of "Chinese boxes" (Simon, 1973) one fitting within the other. This suggestion does not fit the data very well. Consider that the same scientist who is working with far-flung colleagues on an international project may also be working locally with students and may publish with both collaborators. Thus we do not see levels as in a hierarchy, but a continuum of interactions, feedback and exchange, suggesting a heterarchy of partially nested structures (Kontopolous, 1993) that may also be disciplining global connections but not constraining local choices.

Emergence of a hierarchy may be inevitable as groups grow to a certain size (Valverde & Solé, 2007). In systems studies, the system reaches a point of organization where top-down patterns of connection can influence the structure as feedback. In other words, international collaborations— by virtue of attracting attention—become the reference point for everyone in the knowledge system. The 'global' emergences as a hierarchical structure on top of national and disciplinary subsystems. Again, as suggested by Simon (1962), we can begin to describe the properties of the whole: "Hierarchy…is one of the central structural schemes that the architect of complexity uses." Simon suggested that the complex system is composed of subsystems that stabilize and become the scaffold on which the next emergent layer develops. This description fits the data. However, the feedback is not from the top down, as one might expect in a classic hierarchy, but appears to be a series of interactive feedback loops more closely associated with heterarchy than hierarchy.

The structure at the global level does not negate preferential attachment as an evolutionary mechanism at the local level (Jeong et al., 2003; Newman, 2001a) --the growth patterns are consistent with variation and selection in both wider search but also with more elite expectations. A selection mechanism favoring reputation and reward constrains those who collaborate globally because the global level appears to be selective: this can explain the greater numbers of citations to papers at these levels. As Whitley (2001) noted, choice creates competition for connection to the more reputed researchers, an observation which is consistent with these data.

The six cases reveal a global network and specialty networks with resilient and robust structure over time, even as individual nodes enter and leave the network. These findings shed additional light on the underlying dynamic of preferential attachment. Once a network has formed into a resilient structure—in this case, where reputation becomes increasingly important--local interactions may no longer have as much influence on the organization. Padgett and Powell (2012) point out that new organizational forms can emerge in unexpected ways, and transform their environment. Padgett and Powell (2012) also suggest that groups congeal out of iterations of relations, with novelty arising from multiple, intertwined social networks—a concept similar to heterarchy (Kontopolous, 1993). The emergence of novelty may explain the attraction of the international connections. We know that global connections are more likely to be constituted by well-reputed nodes; they are therefore attractive to other actors seeking to enhance their own reputations. Globally connected researchers can, in turn, be highly selective in choosing the next entrant into the network. This enhances the attraction of attaching to other highly reputed nodes at a distance, despite the transaction costs associated with long distance, cross-cultural connections.

Technical Note

1. FWCI is a normalized citation impact measure. It first takes citations per paper, and then divides each paper's citations by average values for field/ publication year/document type. Each paper in the Scopus database has a FWCI score, which changes by year. The value displayed in the table here is the average of those values of the articles in that country-combination for 2013. So if the score is 17.18, it means the article is cited 17.18 times more than the field(s)/year/doctype average. This may seem as a lot, but obviously a single article can be cited 120 times, and still be part of a subject average of 2. Since FWCI is calculated over a large set of articles, its value tends to average back to 1.0, that is, the average FWCI of all articles worldwide.
2. The data for Tables 2 and 4 were drawn from different databases. 1990 and 2000 data were drawn from Web of Science. 2008 and 2013 were drawn from Scopus. The same journals were used in both cases, so we do not expect the different sources to alter the data. In earlier analyses using Pajek, we had specified arcs, which Pajek counted as "2", or, in both directions. Gephi counts an undirected link as "1", or, a connection and discounts isolated nodes. Counting undirected links as a single connection is more common in the literature. As a result, Table 2 shows lower numbers than earlier work (Wagner, 2005).

Acknowledgement

We wish to thank Jeroen Bass at Elsevier for assistance in developing the data for the global level and for the specialties for 2008 and 2013. We thank Meng Li for assistance with statistical analysis.

**Appendix A. List of Journals found to be most central to the specialty, examined for four different years, 1990, 2000, 2008, 2013**

**Astrophysics journals**

*Annual Review of Astronomy and Astrophysics*

*Astrophysical Letters & Communications*

*Astronomy & Astrophysics Supplement Series*

*Astronomy and Astrophysics*

*Astronomical Journal*

*Astronomy Letters-A Journal of Astronomy and Space*

*Astrophysics*

*Astronomy Reports*

*Astrophysical Journal*

*Astrophysical Journal Supplement Series*

*Astrophysics and Space Science*

*Publications of The Astronomical Society of Japan*

*Monthly Notices of the Royal Astronomical Society*

*Publications of The Astronomical Society of the Pacific*

*Solar Physics*

**Mathematical logic**

*Mathematical Logic Quarterly*

*Journal of Symbolic Logic*

*History and Philosophy of Logic*

*Bulletin of Symbolic Logic*

*Archive for Mathematical Logic*

*Annals of Pure and Applied Logic*

**Polymer Science**

*Progress in Polymer Science*

*Polymer Bulletin*

*Macromolecular Symposia*

*Macromolecules*

*Macromolecular Chemistry and Physics*

*Journal of Polymer Science Part A-Polymer Chemistry*

*Journal of Polymer Science Part B-Polymer Physics*

*Journal of Macromolecular Science-Pure and Applied Chemistry*

*European Polymer Journal*

*Biopolymers/Pva Hydrogels/Anionic Polymerisation Nanocomposites*

**Seismology**

*Soil Dynamics and Earthquake Engineering*

*Bulletin of The Seismological Society of America*

*Journal of Seismology*

*Physics of The Earth and Planetary interiors*

*Earth Planets and Space*

*Geophysical Journal international*

*Geophysical Research Letters*

*Journal of Geophysical Research-Solid Earth*

*Tectonophysics*

**Soil Science**

*Advances in Agronomy*

*Australian Journal of Soil Research*

*Canadian Journal of Soil Science*

*Communications in Soil Science and Plant Analysis*

*European Journal of Soil Science*

*Forest Ecology and Management*

*Geoderma*

*Soil Science Society of America Journal*

*Soil Science*

*Soil & Tillage Research*

**Virology**

*Advances in Virus Research*

*Virology*

*Advances in Virus Research*

*Archives of Virology*

*Journal of General Virology*

*Journal of Medical Virology*

*Journal of Virology*

*Journal of Virological Methods*

*Virus Genes*

*Virus Research*

*Journal of Virology*